\documentclass[journal]{IEEEtran}

\usepackage{multirow}
\usepackage[table,xcdraw]{xcolor}
\usepackage{graphicx}
\usepackage{amsmath}
\usepackage{amsfonts}
\usepackage{algorithm}
\usepackage{amsthm,amssymb}
\usepackage{mathrsfs}
\usepackage{algorithmic}
\usepackage{makecell}
\usepackage{easyReview}
\usepackage{booktabs}
\setcellgapes{3pt}
\usepackage{verbatim}
\setcellgapes{3pt}

\usepackage[justification=centering]{caption}
\usepackage{color}
\usepackage[subfigure]{tocloft}
\usepackage{subfigure}
\usepackage{tabularx}
\usepackage{amsmath, bm}
\usepackage{hyperref}
\hypersetup{hidelinks,
	colorlinks=true,
	allcolors=blue,
	pdfstartview=Fit,
	breaklinks=true}

\usepackage{cite}

\ifCLASSINFOpdf
\else
\fi
\begin{document}

%
\title{Large Generative Model Assisted 3D Semantic Communication}

\author{Feibo Jiang, \textit{ Member, IEEE}, Yubo Peng, Li Dong, Kezhi Wang, \textit{Senior Member, IEEE}, Kun Yang, \textit{Fellow, IEEE}, Cunhua Pan, \textit{Senior Member, IEEE}, Xiaohu You, \textit{Fellow, IEEE}
	\thanks{
		Feibo Jiang (jiangfb@hunnu.edu.cn) is with Hunan Provincial Key Laboratory of Intelligent Computing and Language Information Processing, Hunan Normal University, Changsha, China.
		
		Yubo Peng (pengyubo@hunnu.edu.cn) is with School of Information Science and Engineering, Hunan Normal University, Changsha, China.
		
		Li Dong (Dlj2017@hunnu.edu.cn) is with Changsha Social Laboratory of Artificial Intelligence, Hunan University of Technology and Business, Changsha, China.
		
		Kezhi Wang (Kezhi.Wang@brunel.ac.uk) is with the Department of Computer Science, Brunel University London, UK.
		
		Kun Yang (kunyang@essex.ac.uk) is with the School of Computer Science and Electronic Engineering, University of Essex, Colchester, CO4 3SQ, U.K., also with Changchun Institute of Technology.
		
		Cunhua Pan (cpan@seu.edu.cn) is with the National Mobile Communications Research Laboratory, Southeast University, Nanjing 210096, China.
		
		Xiaohu You (xhyu@seu.edu.cn) is with the Frontiers Science Center for Mobile Information Communication and Security, National Mobile Communications Research Laboratory, Southeast University, Nanjing, China, and also with the Purple Mountain Laboratories, Nanjing, China.
	}
}

\markboth{Submitted for Review}%
{Shell \MakeLowercase{\textit{et al.}}: Bare Demo of IEEEtran.cls for IEEE Journals}
%



\maketitle


\begin{abstract}
	Semantic Communication (SC) is a novel paradigm for data transmission in 6G. However, there are several challenges posed when performing SC in 3D scenarios:
	1) 3D semantic extraction;
	2) Latent semantic redundancy; and
	3) Uncertain channel estimation.
	To address these issues, we propose a Generative AI Model assisted 3D SC (GAM-3DSC) system.
	Firstly, we introduce a 3D Semantic Extractor (3DSE), which employs generative AI models, including Segment Anything Model (SAM) and Neural Radiance Field (NeRF), to extract key semantics from a 3D scenario based on user requirements. The extracted 3D semantics are represented as multi-perspective images of the goal-oriented 3D object.
	Then, we present an Adaptive Semantic Compression Model (ASCM) for encoding these multi-perspective images, in which we use a semantic encoder with two output heads to perform semantic encoding and mask redundant semantics in the latent semantic space, respectively.
	Next, we design a conditional Generative adversarial network and Diffusion model aided-Channel Estimation (GDCE) to estimate and refine the Channel State Information (CSI) of physical channels. 
	Finally, simulation results demonstrate the advantages of the proposed GAM-3DSC system in effectively transmitting the goal-oriented 3D scenario.
\end{abstract}

\begin{IEEEkeywords}
	3D Semantic communication;  AIGC; SAM; NeRF; Diffusion Model; GAN.
\end{IEEEkeywords}

\section{Introduction}
In the future, 6G will blur the boundaries between reality and virtuality, reshaping our world to accommodate diverse communication entities, including humans, machines, objects, and spirits. The information exchange between these entities will require higher intelligence, precision, and simplicity \cite{uusitalo20216g}. Therefore, the 6G network will have to facilitate the wireless transmission of huge data volumes, insist on swift system responses, and ensure trustworthy and efficient information interaction \cite{yang2022semantic}.

In response, Semantic Communication (SC) is provided as a viable paradigm for reducing data transmission in 6G \cite{qin2021semantic}. SC dramatically reduces the volume of transmission data by transmitting essential semantic information and reconstructing raw data using Artificial Intelligence (AI) technologies \cite{huang2022towards}. This process can boost efficiency and expand the range and frequency of interactions in multiplayer scenarios in 6G. However, despite these benefits, several challenges remain in implementing SC for 3D scenario transmission:
\begin{enumerate}[]
	\item {\it{3D semantic extraction:}}
	In the future, a wide range of applications, including Augmented Reality (AR) and Mixed Reality (MR), work tirelessly to provide immersive user experiences. As a result, the 3D scenario becomes the primary data type transmitted between these applications. However, the 3D object is represented by voxels or point clouds, which requires transmitting large amounts of data. Few SC models have considered the semantic extraction from 3D scenarios.

	\item {\it{Latent semantic redundancy:}}
	Although semantic encoders can compress information by encoding the source 3D data into a latent semantic space, the Deep Learning (DL) based encoder determines that semantic redundancy still exists in the latent space (i.e., removing certain latent semantic features may not significantly impact the final results) \cite{iyer2022survey}. In future 6G applications, data transmission and exchange occur frequently, hence the redundant semantics contribute to extra communication costs that cannot be overlooked \cite{wang2022semantic}. 
	
	\item {\it{Uncertain channel estimation:}} 
	Complex channels, especially wireless fading channels like Rayleigh and Rician \cite{xiao2006novel}, could impact the recovery of signals and further impact the transmission efficiency of the 3D SC system. Therefore, we should consider high-quality channel estimation technology to ensure the recovery of signals in physical channels.
\end{enumerate}

In recent years, with the rapid advancement of DL, AI has entered the era of content generation, giving rise to numerous novel Generative AI models (GAMs) such as Generative Adversarial Networks (GANs) \cite{goodfellow2020generative}, Diffusion Models (DMs) \cite{yang2022diffusion} and Neural Radiance Field (NeRF) \cite{mildenhall2021nerf}. Leveraging these GAMs and vast amounts of training data, Large AI Models (LAMs) with a significant number of parameters have been applied across various fields. GPT-4 \cite{taloni2023modern} and Segment-Anything Model (SAM) \cite{kirillov2023segment} are both typical representatives of LAMs.
These LAMs offer several advantages for SC, such as accurate semantic extraction, rich prior/background knowledge, and robust semantic interpretation \cite{jiang2023large,jiang2023large3}.
Therefore, to address the aforementioned challenges, we propose a GAM assisted 3D SC (GAM-3DSC) system, in which we consider the semantic extraction of 3D scenarios, semantics compression to reduce redundancy, and high-quality channel estimation. The main contributions are summarized as follows:
\begin{enumerate}[]
	\item 
	We introduce a 3D Semantic Extractor (3DSE) to perform the goal-oriented semantic extraction of a 3D scenario. Specifically, first, we utilize the User Equipment (UE) to obtain the images of a 3D scenario from different perspectives.
	Then, we apply SAM to allow users to select the key 3D object from the captured images. Next, we utilize mask inverse rendering technology to obtain the multi-perspective images of the selected 3D object. These multi-perspective images can be viewed as the semantics of the raw 3D scenario. Following this, we employ an image-based SC model to transmit these multi-perspective images. Finally, in the receiver, we employ NeRF to construct the 3D scenario based on received multi-perspective images.
	
	\item We apply an Adaptive Semantic Compression Model (ASCM) to achieve multi-perspective image SC. Specifically, we present a semantic encoder with two output heads to carry out encoding while simultaneously eliminating redundant semantic information in the latent feature space, thereby achieving semantic compression. Furthermore, during the training phase, we employ a Self-Knowledge Distillation (SKD) to direct the semantic compression, minimizing the discrepancy between final decoded results using semantic compression and not. As a result, we remove the redundant semantics and reduce the communication cost.
	
	\item We develop a Conditional Generative Adversarial Network (CGAN) and Diffusion Model (DM) aided-Channel Estimation (GDCE). Therefore, we use CGAN to estimate the Channel State Information (CSI) of physical channels, where the pilot sequence is conditional information. We then employ a DM to further refine the estimated CSI. With the acquired CSI, 
	we enhance the recovery of signals in physical channels.
\end{enumerate}

The structure of this paper is as follows. Section II introduces the preliminaries. Section III provides a detailed description of the SC system model and problem description. 
Section IV presents the proposed GAM-3DSC system, which includes the 3DSE, ASCM, and GDCE schemes. 
Section V employs numerical results to evaluate the performance of the proposed GAM-3DSC. Lastly, Section VI concludes this paper.

\section{Preliminaries}
This section introduces the preliminaries about the key GAMs used in this paper, including NeRF, SAM, CGAN, and DM.

\subsection{Neural Radiance Field}
NeRF is a method that leverages DL to extract geometric shape and texture information of objects from images taken from multiple perspectives. This information is then used to generate a continuous three-dimensional radiance field, enabling the reconstruction of highly realistic three-dimensional models at any angle and distance using a few multi-perspective images \cite{mildenhall2021nerf}.
The key idea of NeRF involves the implicit learning of a static 3D scenario using a Multilayer Perceptron (MLP). Specifically, given the 3D coordinate position and viewing direction of a spatial point as input, MLP can output both color and density for that point. 
As an innovative method for perspective synthesis and 3D reconstruction, NeRF holds promising application potential in diverse fields such as robotics, urban mapping, autonomous navigation, and VR/AR among others. 
In \cite{guo2022nerfren}, the author presented NeRFReN, an extension of NeRF designed specifically to model scenarios that involve reflections, enabling accurate modeling and representation of scenarios with reflective surfaces. 
In \cite{maggio2023loc}, the author introduced Loc-NeRF, an innovative real-time vision-based robot localization method that combined the strengths of Monte Carlo localization and NeRF, achieving accurate and efficient robot localization in real-world environments. 

\subsection{Segment-Anything Model}
SAM, introduced by Meta AI, represents a groundbreaking segmentation system for images \cite{kirillov2023segment}. SAM is trained on the largest and most diverse dataset to date, the Segment Anything 1-Billion (SA-1B). This dataset includes over 1 billion masks across 11 million licensed and privacy-conscious images \cite{kirillov2023segment}.
SAM applies an efficient transformer-based architecture that is adept at both natural language processing and image recognition tasks. This architecture is composed of a visual transformer-based image encoder for feature extraction, a prompt encoder for user interaction, and a mask decoder for generating segmentation and confidence scores.
This innovative system can effectively carry out zero-shot segmentation for previously unseen images or objectives without necessitating further knowledge or training.
In \cite{ma2023segment}, the authors introduced MedSAM, the pioneering foundation model specifically designed for comprehensive medical image segmentation.
Building on the SAM model, \cite{yu2023inpaint} represented the first effort towards mask-free image inpainting, introducing a novel paradigm termed ``clicking and filling". 
In \cite{tang2023can}, the author explored whether SAM can effectively tackle the challenging task of Camouflaged Object Detection (COD).

\subsection{Conditional Generative Adversarial Network}
A Generative Adversarial Network (GAN) is a deep learning model composed of a generator and a discriminator, trained concurrently in a game-like scenario. This approach enables the generator to produce realistic data while the discriminator strives to differentiate between generated and real data \cite{goodfellow2020generative}. 
CGAN merges the principles of GANs and conditional generation, introducing conditional information. This allows the generative model to create data under specific conditions rather than merely generating random samples \cite{zhang2019image}. 
CGAN has found widespread use in various generative tasks such as image creation, image restoration, style transfer, and generative adversarial network generation.
In \cite{zhang2021distributed}, the authors proposed a highly effective channel estimation approach that allows each drone to train a dedicated channel model independently for each beamforming direction using a CGAN.
The author in \cite{9837473} introduced a groundbreaking approach using CGANs for uplink-to-downlink mapping of both channel covariance matrices (CCMs) and CSI. \cite{10225927} investigated an adversarial learning-based approach for wireless signal denoising, in which the author designed a CGAN at the receiver to establish an adversarial game between a generator and a discriminator.

\subsection{Diffusion Model}
The DM \cite{yang2022diffusion} is an advanced generative model that excels at producing high-quality data by employing a gradual ``diffusion" process to eliminate image noise. 
DM showcases remarkable capabilities in generating progressively more realistic and diverse samples, through leveraging progressive training techniques, autoregressive generation methods, and potentially combining with GANs. 
DM has also demonstrated exceptional performance in various domains such as image generation and other challenging generative tasks, making it a highly versatile and effective tool in LAMs.
In \cite{saharia2022palette}, the authors developed a unified framework for image-to-image translation based on conditional DMs and evaluated this framework on four challenging image-to-image translation tasks. 
To generate long and higher resolution videos, reference \cite{ho2022video} introduced a new conditional sampling technique for spatial and temporal video extension. 
In \cite{xu2022geodiff}, the authors proposed a novel generative model for molecular conformation prediction, where they approached each atom as a particle and reversed the diffusion process directly as a Markov chain.

\section{System Model}
\begin{figure}[htbp]
	\centering
	\includegraphics[width=8.5cm]{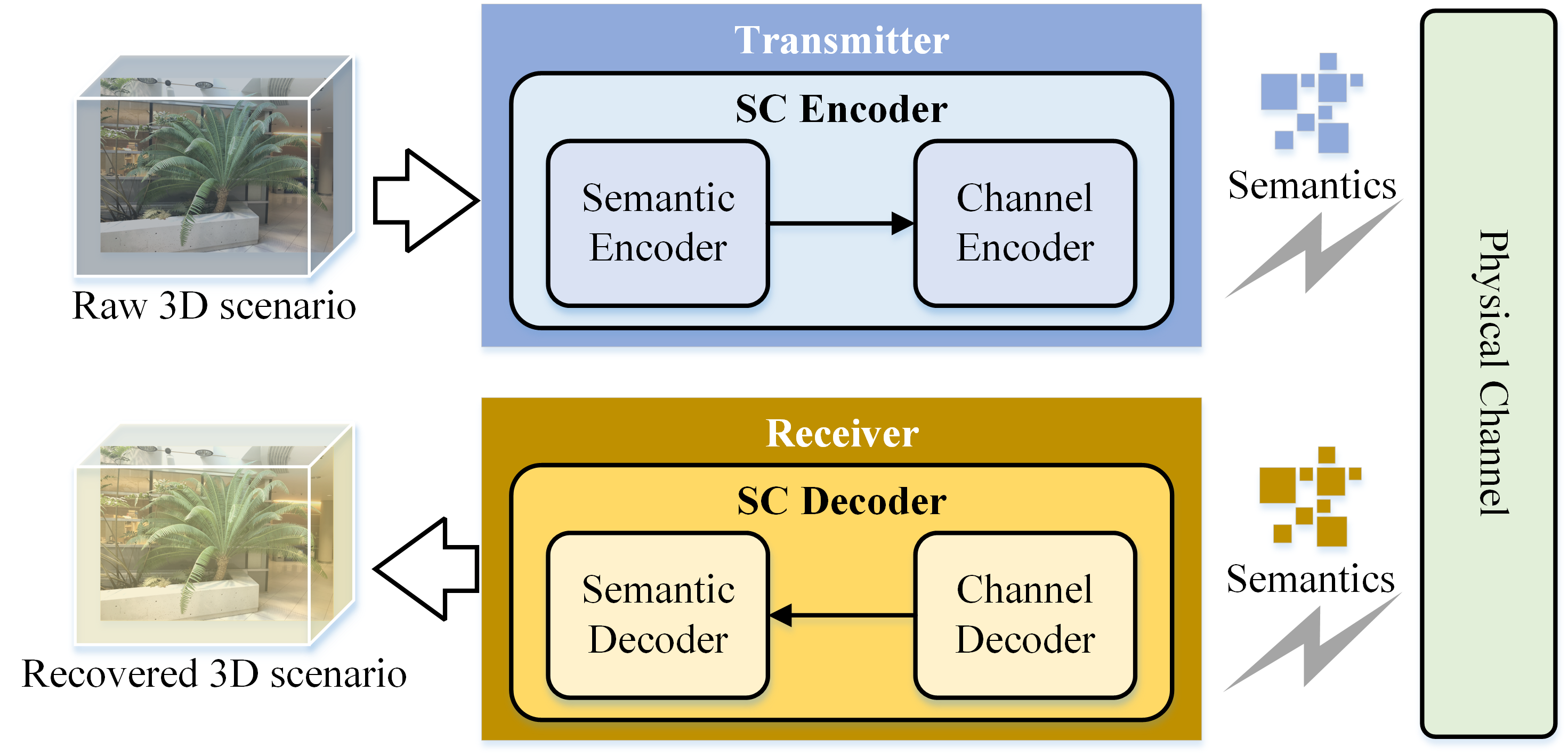}
	\caption{The illustration of the 3D SC between a transmitter and a receiver.}
	\label{fig:system}
\end{figure}

As shown in Fig. \ref{fig:system}, we consider a 3D scenario where a transmitter performs data transmission with a receiver using SC in wireless networks. The transmitter only needs to transmit the semantic information of the raw 3D scenario to the receiver. With the received semantic information, the receiver can recover and obtain the 3D scenario. 
To enable the transmitter can extract the semantics from the raw 3D scenario, we deploy the SC encoder on it. We also deploy the SC decoder on the receiver to perform semantic decoding, recovering the 3D scenario according to the received semantics.

The performance of SC mainly depends on the encoding and decoding of the semantic information, hence the architecture design of the SC model must be reasonable and suitable.
The SC encoder includes the semantic and channel encoders, extracting the semantics from the raw 3D scenario $\chi^\text{3D}$. Then, the semantics are encoded and modulated to be able to transmit over a wireless channel. The result processed by the SC encoder can be expressed as:
\begin{equation}\label{eq:Shi1}
	\mathbf{X} = C_\alpha\left(S_\vartheta\left(\chi^\text{3D}\right)\right),
\end{equation}
where $\mathbf{X}$ represents the encoded symbol stream; $S_\vartheta\left(\cdot\right)$ represents the semantic encoder with model parameters $\mathbf{\vartheta}$ and $C_\alpha\left(\cdot\right)$ is the channel encoder with model parameters $\mathbf{\alpha}$. 

When transmitted over the wireless fading channel, $\mathbf{X}$ suffers transmission impairments that include distortion and noise. This transmission process can be given by:
\begin{equation}\label{eq:Shi2}
	\mathbf{Y} = \mathbf{HX+N},
\end{equation}
where $\mathbf{Y}$ represents the received signal; $\mathbf{H}$ represents the channel gain between the transmitter and the receiver; $\mathbf{N}$ is the Additive White Gaussian Noise (AWGN). For end-to-end training of the encoder and decoder, the transmission channel must allow backpropagation. Therefore, we can simulate the channel by neural networks \cite{park2020end}. 

The SC decoder includes the channel and semantic decoders, decoding the received $\mathbf{Y}$ and alleviating transmission impairments. The recovered 3D scenario $\hat{\chi}^\text{3D}$ can be given by:
\begin{equation}\label{eq:Shi3}
	\hat{\chi}^\text{3D}=S^{\prime}_\delta\left(C^{\prime}_\beta(\mathbf{Y})\right),
\end{equation}
where $C^{\prime}_\beta\left(\cdot\right)$ represents the channel decoder with model parameters $\mathbf{\beta}$; $S^{\prime}_\delta\left(\cdot\right)$ is the semantic decoder with model parameters $\mathbf{\delta}$.

\section{Proposed GAM-3DSC System}
Since implementing SC in the transmission of a 3D scenario presents three critical challenges: 3D semantic extraction, latent semantic redundancy, and uncertain channel estimation, we propose the GAM-3DSC system as a solution in this paper. In GAM-3DSC, we first design the 3DSE to perform semantic extraction from the raw 3D scenario, by utilizing the strength of NeRF and SAM. We detail 3DSE in Section IV-B. 
Then, we present the ASCM to realize the multi-perspective image SC between the transmitter and receiver. ASCM could compress the semantics by removing unimportant information, thus reducing communication costs efficiently. We detail ASCM in Section IV-C. 
Finally, we apply GDCE to perform the channel estimation, where we use CGANs to obtain the CSI of the channel and then utilize a DM to refine CSI, thus enhancing the recovery of received signals. We detail GDCE in Section IV-D. 

\subsection{Overview}
\begin{figure*}[htbp]
	\centering
	\includegraphics[width=17cm]{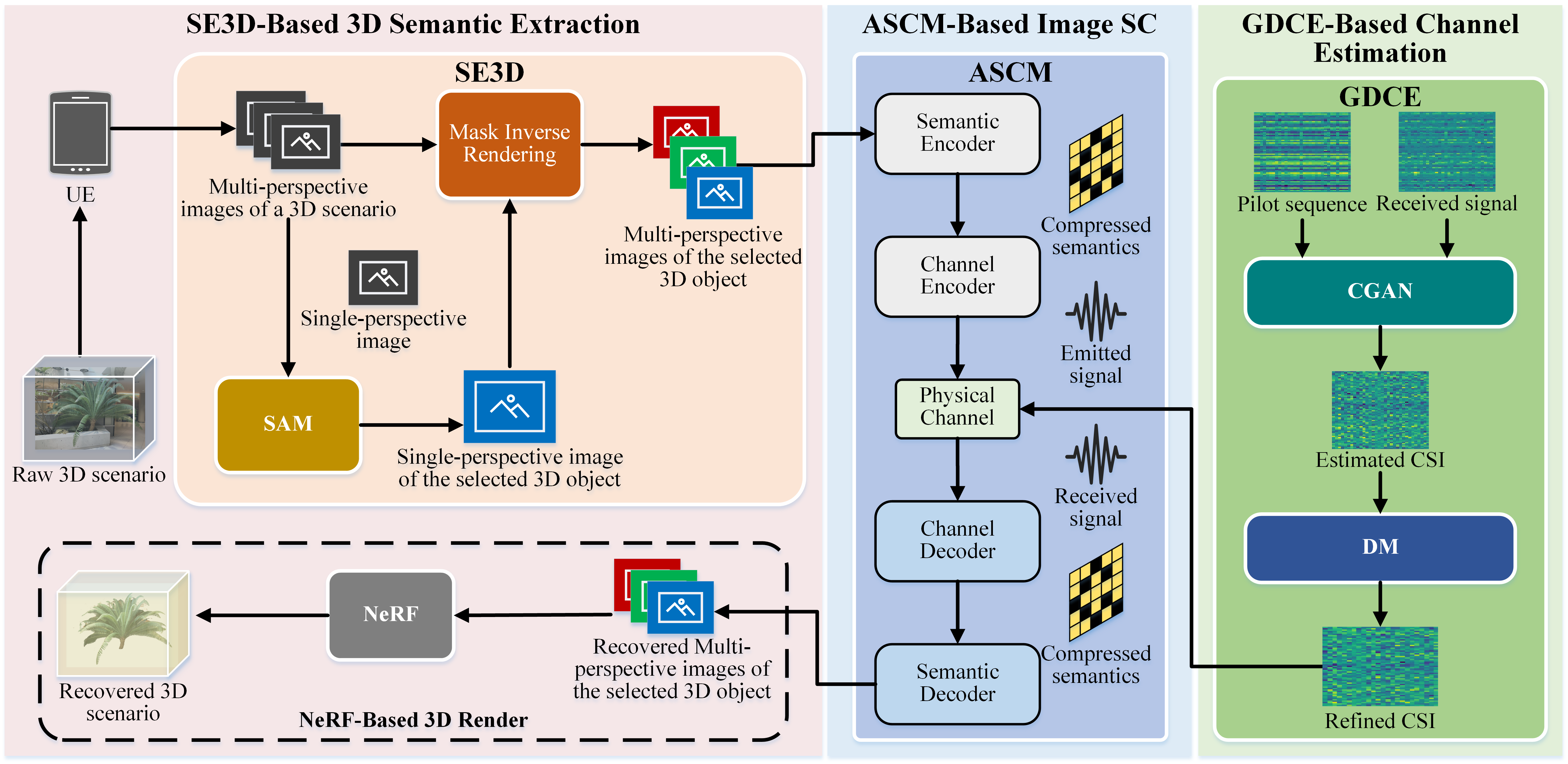}
	\caption{The illustration of the proposed GAM-3DSC system.}
	\label{fig:LAM-3DSC}
\end{figure*}
In this subsection, we introduce the workflow of the proposed GAM-3DSC system. As shown in Fig. \ref{fig:LAM-3DSC}, the GAM-3DSC is divided into three parts, including the 3DSE-based 3D semantic extraction, ASCM-based efficient image SC, and GDCE-based channel estimation. Next, we introduce the three parts, respectively. 

\subsubsection{3DSE-Based Semantic Extraction}
First, in the transmitter, we use the UE to capture images $\mathbf{I}_\text{M}$ of a 3D scenario $\chi^\text{3D}$ from various perspectives. Then, we apply SAM to enable users to select a key 3D object from one of the multi-perspective images $\mathbf{I}_\text{M}$ according to their specific goals. Subsequently, we employ mask inverse rendering technology to obtain multi-perspective images of the selected 3D object, denoted as $\mathbf{I}_\text{M}^\prime$, which we can consider as the semantics of the raw 3D scenario.
This process of semantic extraction is illustrated in \textbf{Algorithm \ref{alg:3DSE}}.
Finally, at the receiver end, we leverage NeRF to construct the 3D scenario based on the recovered multi-perspective images of the goal-oriented 3D object in $\hat{\chi}^\text{3D}$.

\subsubsection{ASCM-Based Image SC}
For the multi-perspective images of the goal-oriented 3D object, we utilize the ASCM to transmit them one by one. The training and inference process of ASCM is described in \textbf{Algorithm \ref{alg:ASCM}}.
With the semantic encoder of ASCM, the images are transformed into semantic information in the latent feature space.
However, DL-based semantic encoders can only reduce pixel-level redundancy and are unable to reduce feature-level redundancy.
Hence, our semantic encoder also outputs a mask array to mask redundant feature-level data in the semantic information. 
This process reduces the volume of the transmitted data, consequently reducing communication costs.

\subsubsection{GDCE-Based Channel Estimation}
Before executing ASCM, we feed a smaller pilot sequence into the trained CGAN, thus we can estimate the CSI of the physical channel.
Due to the prior constraints on CGAN, the generated CSI images often lack fine details. Therefore, an additional DM is employed to enhance the predicted details of CSI.
The training and inference steps of GDCE are described in \textbf{Algorithm \ref{alg:GDCE}}. This approach helps us better perform the recovery of signals in the receiver.

For better understanding, we summarize the proposed GAM-3DSC system in \textbf{Algorithm \ref{alg:LAM-3DSC}}.

\begin{algorithm}
	\caption{GAM-3DSC}
	\label{alg:LAM-3DSC}
	\begin{algorithmic}[1]
		\REQUIRE $\chi^\text{3D}$.
		\ENSURE $\hat{\chi}^\text{3D}$.
		\STATE{Select a key 3D object in $\chi^\text{3D}$ and obtain multi-perspective images $\mathbf{I}_\text{M}^\prime$ of the selected 3D object according to the inference stage of \textbf{Algorithm \ref{alg:3DSE}}.}
		\STATE{Obtain CSI based on GDCE according to the inference stage of \textbf{Algorithm \ref{alg:GDCE}}.}
		\FOR{each single-perspective image in $\mathbf{I}_\text{M}^\prime$}
		\STATE{Apply ASCM and refined CSI to perform efficient image transmission according to the inference stage of \textbf{Algorithm \ref{alg:ASCM}}.}
		\ENDFOR
		\STATE{Recover the 3D scenario $\hat{\chi}^\text{3D}$ using NeRF based the received multi-perspective images.}
	\end{algorithmic}
\end{algorithm}
\subsection{3DSE}
\begin{figure}[htbp]
	\centering
	\includegraphics[width=8.5cm]{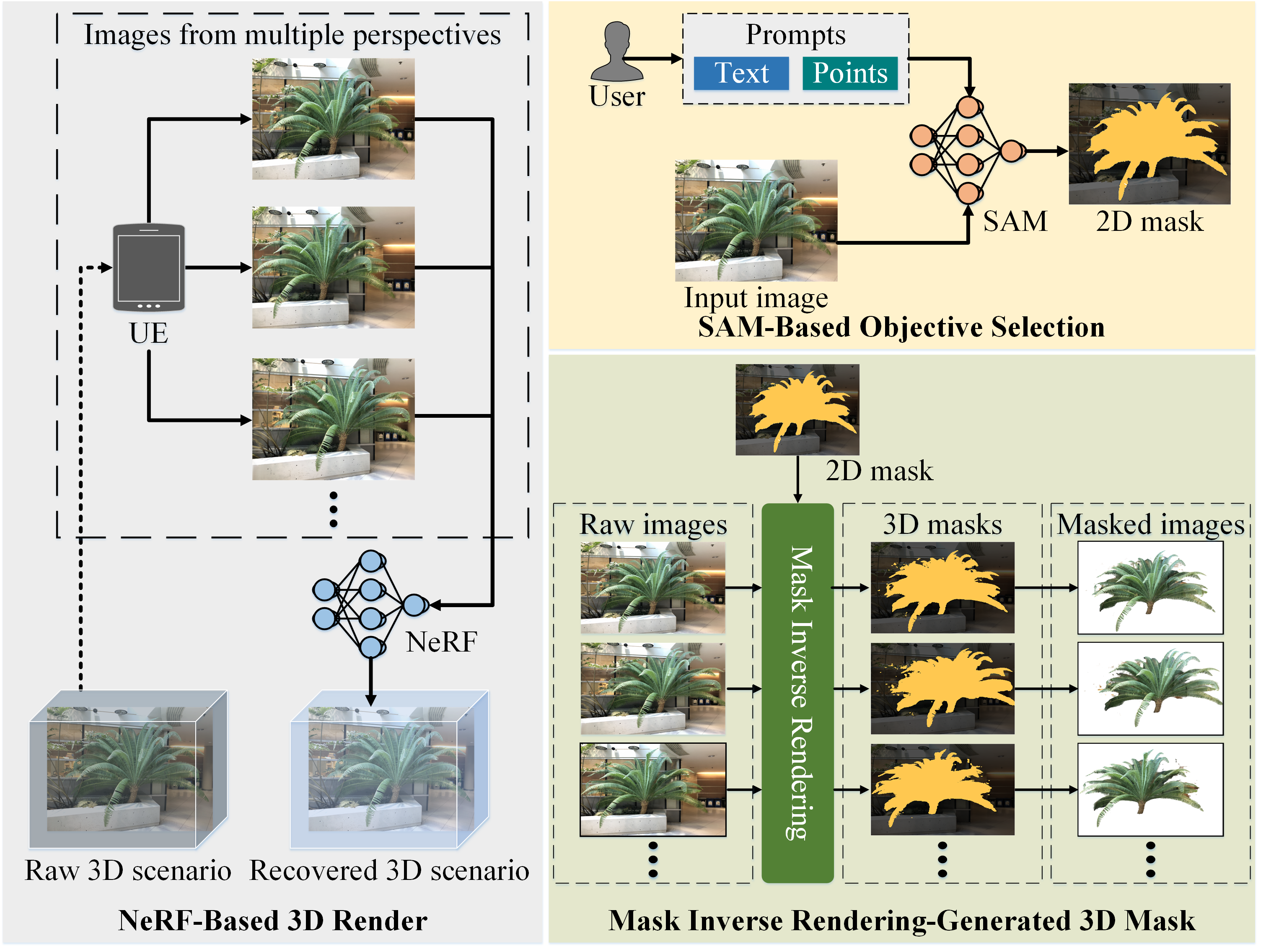}
	\caption{The architecture of 3DSE.}
	\label{fig:3DSE}
\end{figure}

In this subsection, we detail 3DSE, which introduces SAM and NeRF as tools for achieving 3D semantic extraction. As illustrated in Fig. \ref{fig:3DSE}, the components of the presented 3DSE is as follows:

\subsubsection{NeRF-Based 3D Render}
Firstly, we capture images $\mathbf{I}_\text{M}$ from multiple perspectives of a static 3D scenario $\chi^\text{3D}$ using UE. 
Subsequently, we define a neural network $F_ {\xi}$ as:
\begin{equation}\label{eq:3DSE1}
	F_ {\xi}: (\mathbf {x}, \mathbf {d}) \rightarrow (\mathbf {c}, \nu),
\end{equation}
where $\mathbf {x} = (x,y,z)$ signifies the coordinates of a point in three-dimensional space, $\mathbf {d}$ represents the observation direction (expressed in spherical coordinates), $\mathbf {c}$ indicates the color (expressed as RGB values), and $\nu$ denotes the density (expressed as a scalar).

Next, for each pixel in every image, we compute a ray $\mathbf {r} (t)$ originating from the camera based on the camera parameters:
\begin{equation}\label{eq:3DSE2}
	\mathbf {r} (t) = \mathbf {o} + t\mathbf {d},
\end{equation}
where $\mathbf {o}$ denotes the origin of the light (i.e., the position of the camera), $t$ is the parameter on the ray (expressed as a scalar), and $\mathbf {d}$ signifies the direction of the ray (i.e., the direction corresponding to the pixel). Following this, we sample several points on the ray and input these points along with their corresponding directions into the neural network $F_ {\xi}$ to obtain predictions of color $\mathbf {c} (\mathbf {r} (t),\mathbf {d})$ and density $\nu (\mathbf {r} (t))$.

Subsequently, based on the color and density predictions, we calculate the final color $C (\mathbf {r})$ of the ray via volume rendering:
\begin{equation}\label{eq:3DSE3}
	C (\mathbf {r}) = \int_ {t_n}^{t_f} T (t)\cdot\nu (\mathbf {r} (t))\cdot\mathbf {c} (\mathbf {r} (t),\mathbf {d})dt,
\end{equation}
where $T (t)=\exp (-\int_ {t_n}^t \nu (\mathbf {r} (s))ds)$ can be interpreted as the probability that the light traveling from $t_n$ to $t$ is not obstructed by any object. $t_n$ and $t_f$ denote the near and far bounds of the ray, respectively.

Lastly, we calculate the loss function by comparing this value with the actual color of the image:
\begin{equation}\label{eq:3DSE4}
	\mathcal{L}(\xi) = \sum_{m=1}^M \sum_{p=1}^P ||C_m(\mathbf{r}_p) - \hat{C}_m(\mathbf{r}_p)||^2,
\end{equation}
where $M$ represents the number of images; $P$ denotes the number of pixels in each image; $C_m(\mathbf{r}_p)$ is the true color of the $p$-th pixel on the $m$-th image; $\hat{C}_m(\mathbf{r}_p)$ is the predicted color of the $p$-th pixel on the $m$-th image.
By implementing optimization methods such as backpropagation and the Stochastic Gradient Descent (SGD) algorithm \cite{zhang2018improved}, we can update the parameters $\xi$ of the neural network $F_ {\xi}$ to minimize the loss function.

\subsubsection{SAM-Based Objective Selection}
To allow users to freely select a 3D object of interest from the 3D scenario $\chi^\text{3D}$, we use SAM to perform semantic segmentation on the image from a specific perspective. Next, we will introduce the workflow of SAM \cite{kirillov2023segment,jiang2023large2}.

Firstly, we provide a specific single-perspective image $\mathbf{I}_\text{S} \in \mathbf{I}_\text{M}$ as input, along with a prompt $\mathbf{P}$. This prompt can be either text or 2D points and used to identify the objectives for segmentation. When the prompt is in text, we need to convert it into points. For example, if the text prompt is ``flower", the conversion result would be the center point coordinates of the flower.

Subsequently, we feed the input image $\mathbf{I}_\text{S}$ and prompts $\mathbf{P}$ into a neural network $F_{\Gamma}$:
\begin{equation}\label{eq:3DSE5}
	F_{\Gamma}: (\mathbf{I}_\text{S}, \mathbf{P}) \rightarrow (\mathbf{M}, \mathbf{S}, \mathbf {L}),
\end{equation}
where $\mathbf{M}$ refers to the generated mask with a shape of $(H, W)$. Each element of $\mathbf{M}$ is either 0 or 1, indicating whether the pixel is part of the target object or not. $\mathbf{S}$ is the Intersection over Union (IoU) score, which depicts the intersection ratio between the mask and the actual annotation. $\mathbf{L}$ is a category label, denoting the category of the target object. The predicted category label $\hat{\mathbf{L}}$ can be computed as follows:
\begin{equation}\label{eq:3DSE6}
	\hat{\mathbf{L}} = \zeta(\hat{\mathbf{M}}, \hat{\mathbf{S}}, \mathbf{P}),
\end{equation}
where $\zeta(\cdot)$ denotes a function that generates class label predictions based on mask predictions $\hat{\mathbf{M}}$, IoU score predictions $\hat{\mathbf{S}}$, and prompt $\mathbf{P}$.

Finally, after the training process, SAM is capable of generating precise mask predictions.

\subsubsection{Mask Inverse Rendering-Generated 3D Mask}
Since the mask generated by SAM is 2D, we utilize mask inverse rendering to project this 2D mask into 3D space, resulting in a 3D mask. Note the 3D mask refers to the masks of the images from all perspectives.
Technically, we represent the 3D mask as voxel grids $V \in \mathbb{R}^{L\times W \times H}$, where each grid vertex maintains a zero-initialized soft mask confidence score. Using these voxel grids, we render each pixel of the 2D mask from a specific single perspective \cite{cen2023segment}:
\begin{equation}\label{eq:3DSE7}
	\mathbf{M}_\text{3D}(\mathbf {r}) = \int_ {t_n}^{t_f} \omega (\mathbf {r} (t))\cdot\mathbf {U} (\mathbf {r} (t))dt,
\end{equation}
where $\omega (\mathbf {r} (t))$ is derived from the density values of the pre-trained NeRF; $\mathbf {U} (\mathbf {r} (t))$ signifies the mask confidence score at location $\mathbf {r} (t)$, obtained from voxel grids $\mathbf {U}^2$. 
We denote $\mathbf{M}_\text{SAM}(\mathbf{r})$ as the corresponding mask generated by SAM under ray $\mathbf{r}$. When $\mathbf{M}_\text{SAM}(\mathbf{r})=1$, the objective of mask inverse rendering is to enhance $\mathbf {U} (\mathbf {r} (t))$ in relation to $\omega (\mathbf {r} (t))$. 
In practice, we can optimize this by using the SGD algorithm. To serve this purpose, we define the mask projection loss as the negative product between $\mathbf{M}_\text{SAM}(\mathbf{r})$ and $\mathbf{M}_\text{3D}(\mathbf {r})$:
\begin{equation}\label{eq:3DSE8}
	\mathcal{L}_{\text {proj }}=-\sum_{\mathbf{r} \in \mathcal{R}(\mathbf{I}_\text{S})} \mathbf{M}_\text{SAM}(\mathbf{r}) \cdot \mathbf{M}_\text{3D}(\mathbf{r}),
\end{equation}
where $\mathcal{R}(\mathbf{I}_\text{S})$ denotes the ray set of the image $\mathbf{I}_\text{S}$. 

We formulate the mask projection loss based on the assumption that both the geometry from the NeRF and the segmentation results of SAM are accurate. However, in real-world situations, this is not always the case. As a result, we add a negative refinement term to the loss function to optimize the 3D mask grids in line with multi-perspective mask consistency:
\begin{equation}\label{eq:3DSE9}
\begin{split}
	&\mathcal{L}_{\text {proj }}=-\sum_{\mathbf{r} \in \mathcal{R}(\mathbf{I}_\text{S})} \mathbf{M}_{\mathrm{SAM}}(\mathbf{r}) \cdot  \mathbf{M}_\text{3D}(\mathbf{r})+\\
	&\lambda \sum_{\mathbf{r} \in \mathcal{R}(\mathbf{I}_\text{S})}\left(1-\mathbf{M}_{\mathrm{SAM}}(\mathbf{r})\right) \cdot \mathbf{M}_\text{3D}(\mathbf{r}),
\end{split}
\end{equation}
where $\lambda$ serves as a hyperparameter to govern the magnitude of the negative term. In each iteration, we update the 3D mask $\mathbf{U}$ by employing the SGD algorithm to minimize $\mathcal{L}_{\text {proj }}$. 
After obtaining the 3D mask $\mathbf{U}$, we multiply it with $\mathbf{I}_\text{M}$ to obtain $\mathbf{I}_\text{M}^\prime$:
\begin{equation}\label{eq:3DSE10}
	\mathbf{I}_\text{M}^\prime=\mathbf{I}_\text{M}\odot \mathbf{U}.
\end{equation} 

For SAM, we can utilize the pre-trained weight presented in \cite{kirillov2023segment}, thus we only need to train the NeRF in 3DSE. We summarize the 3DSE approach in \textbf{Algorithm \ref{alg:3DSE}}.
\begin{algorithm}
	\caption{3DSE}
	\label{alg:3DSE}
	\begin{algorithmic}[1]
		\REQUIRE $\chi^\text{3D}$.
		\ENSURE $\mathbf{I}_\text{M}^\prime$.
		\STATE{\textbf{Training Stage}}
		\STATE{Acquire multi-perspective images of $\chi^\text{3D}$ using the camera of UE in different positions.}
		\STATE{Calculate the loss function $\mathcal{L}(\xi)$ of NeRF in accordance with Eqs. (\ref{eq:3DSE1})-(\ref{eq:3DSE4}).}
		\STATE{Update $\xi$ by applying the SGD algorithm to minimize $\mathcal{L}(\xi)$.}
		\STATE{\textbf{Inference Stage}}
		\STATE{Input an image $\mathbf{I}_\text{S}$ from a specific single perspective into SAM to yield the 2D mask $\mathbf{M}_\text{SAM}$.}
		\STATE{Perform reverse rendering from 2D mask to obtain 3D mask based on Eq. (\ref{eq:3DSE7}).}
		\STATE{Calculate the loss function $\mathcal{L}_\text{proj}$ according to Eq. (\ref{eq:3DSE9}).}
		\STATE{Iteratively update 3D mask $\mathbf{U}$ by applying the SGD algorithm to minimize $\mathcal{L}_\text{proj}$.}
		\STATE{Obtain the multi-perspective images $\mathbf{I}_\text{M}^\prime$ of the selected 3D object according to Eq. (\ref{eq:3DSE10})}.
	\end{algorithmic}
\end{algorithm}

In conclusion, there are two advantages to 3DSE: (1) We employ NeRF to render the 3D scenario, requiring only a few images from different perspectives. This implies that we merely transmit some 2D images when transferring 3D scenario data, thereby reducing the cost and complexity of communications. (2) We utilize SAM to allow users to select a 3D object of interest in the 3D scenario from a specific perspective image, thereby attaining accurate semantic extraction that aligns with human perception.

\subsection{ASCM}
\begin{figure}[htbp]
	\centering
	\includegraphics[width=8.5cm]{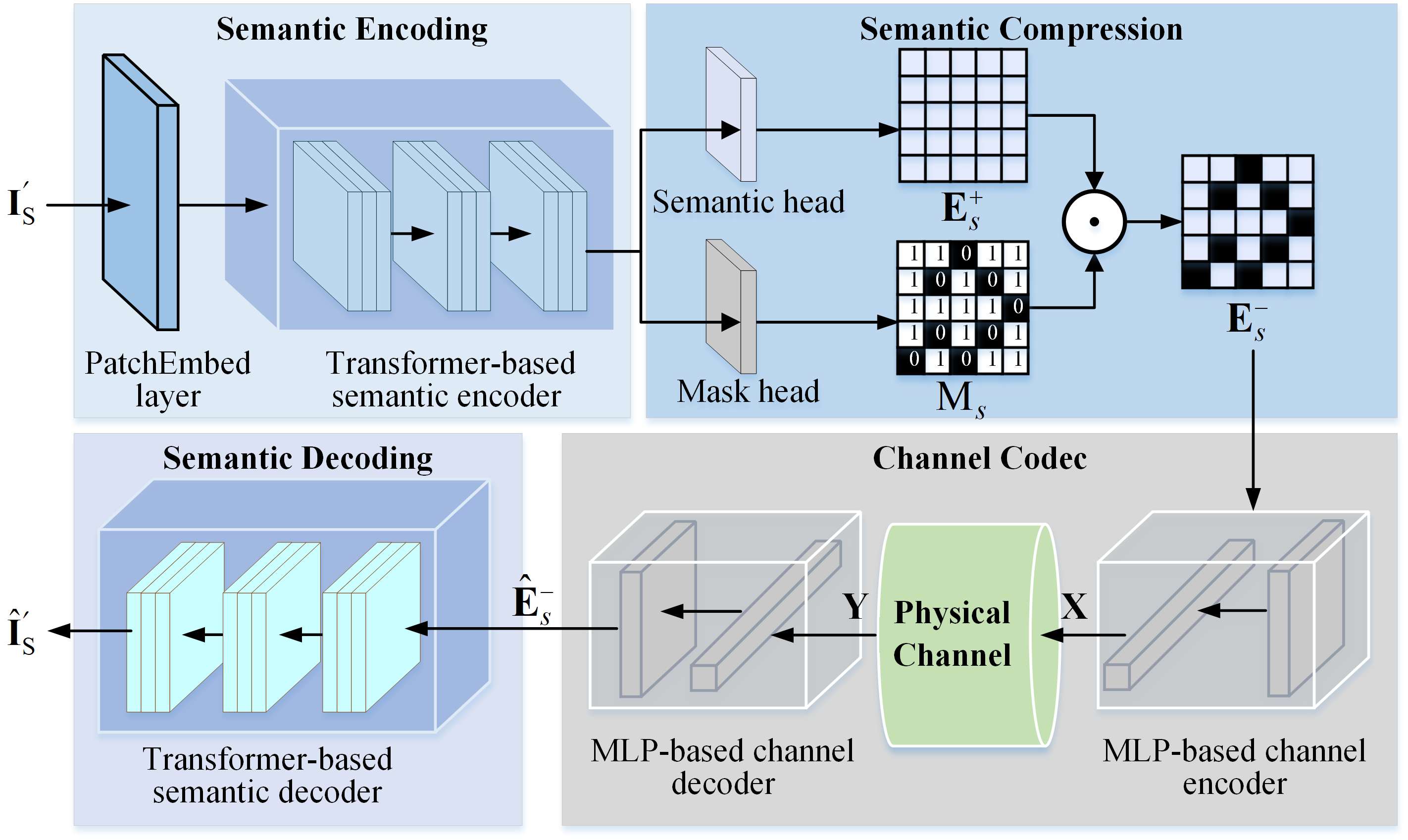}
	\caption{The architecture of ASCM.}
	\label{fig:ASCM1}
\end{figure}
\begin{figure}[htbp]
	\centering
	\includegraphics[width=8.5cm]{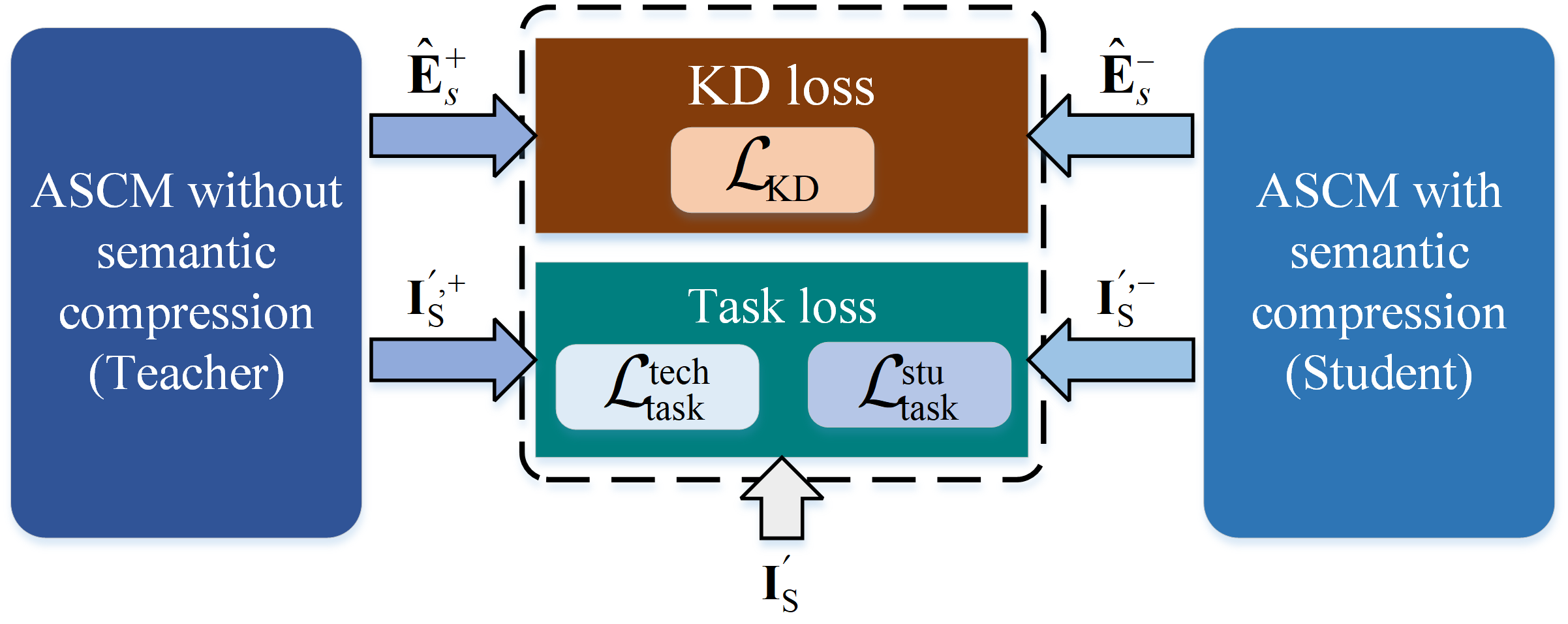}
	\caption{SKD-based model training for ASCM.}
	\label{fig:ASCM2}
\end{figure}
In this subsection, we introduce ASCM to optimize image SC between transmitters and receivers. 
ASCM can learn to compress the feature-level redundancy in semantics derived from raw image data, thereby minimizing the volume of transmitted semantic information. As depicted in Fig. \ref{fig:ASCM1}, the workflow of the presented ASCM is described as follows:

\subsubsection{Semantic Encoding}
To extract the semantics from the single-perspective image $\mathbf{I}_\text{S}^\prime$ ($\in \mathbf{I}_\text{M}^\prime$) of the selected 3D object, we employ a DL network as the semantic encoder of ASCM. Transformer networks \cite{khan2022transformers} have proven more effective than traditional Convolutional Neural Networks (CNNs), due to their superior performance in various domains such as computer vision, natural language processing, and speech processing. Therefore, we utilize the transformer networks as the semantic encoder. 
Firstly, we convert $\mathbf{I}_\text{S}^\prime$ into a sequence of one-dimensional patch embeddings via a PatchEmbed layer \cite{zhou2023speech}. Then, the transformer networks extract semantic features from the embeddings. The transformer networks consist of multiple stacked attention layers and utilize the following attention formula:
\begin{equation}\label{eq:ASCM1}
	\operatorname{Attention}(\mathbf{Q}, \mathbf{K}, \mathbf{V})=\operatorname{Softmax}\left(\frac{\mathbf{Q} \mathbf{K}^\top}{\sqrt{d}}\right) \mathbf{V},
\end{equation}
where $\mathbf{Q}$, $\mathbf{K}$, and $\mathbf{V}$ are all obtained by linear transformation of the input embeddings; $d$
is the adjustment factor; $\operatorname{Softmax}(\cdot)$ is an activation function.   

\subsubsection{Semantic Compression}
To further reduce communication costs, we propose eliminating redundant data in the semantic features. 
In the semantic encoder, we design two output heads, called semantic head and mask head. 
The semantic head output the full semantics $\mathbf{E}_s^+$, and the mask head output a mask array $\mathbf{M}_s=[m_1,m_2,..,m_i,...,m_S]$, where $S$ signifies the length of $\mathbf{E}_s^+$. Here, either $m_i=0$ or $m_i=1$, and $m_i$ determines whether each feature in $\mathbf{E}_s^+$ is reserved. When the $i$-th semantic feature in $\mathbf{E}_s^+$ is retained, then $m_i=1$ otherwise $m_i=0$. Thus, the compressed semantic information $\mathbf{E}_s^{-}=\mathbf{E}_s^+\odot\mathbf{M}_s$.

\subsubsection{Channel Codec}
To enable semantic information to be transmitted over wireless channels, we construct the channel encoder and decoder based on the MLP networks. 
The channel encoder is used to modulate and encode $\mathbf{E}_s^{-}$. Then, the result of the encoded semantic information can be given by:
\begin{equation}\label{eq:ASCM3}
	\mathbf{X}=\sigma\left(\mathbf{E}_s^{-} \mathbf{w}_{t}+\mathbf{b}_{t}\right),
\end{equation}
where $\sigma(\cdot)$ represents an activation function; $\mathbf{w}_t$ is the weight matrix of the channel encoder; $\mathbf{b}_t$ is the bias.

According to Eq. (\ref{eq:Shi2}), after transmission on the wireless channel, $\mathbf{X}$ changes into $\mathbf{Y}$. The channel decoder performs demodulating $\mathbf{Y}$ and removes the impairments from the channel. Then, the decoding results of the channel decoder can be expressed as:
\begin{equation}\label{eq:ASCM4}
	\hat{\mathbf{E}}_{s}^{-} =
	 \sigma\left(\mathbf{Y} \mathbf{w}_{r}+\mathbf{b}_{r}\right),
\end{equation}
where  $\mathbf{w}_r$ is the weight matrix of the channel decoder; $\mathbf{b}_r$ represent the bias.

\subsubsection{Semantic Decoding}
The semantic decoder performs decoding the received semantics $\hat{\mathbf{E}}_{s}^{-}$, where the semantic decoder is also constructed based on the transformer networks. The decoded result is the reconstructed single-perspective image $\mathbf{\hat{I}}_\text{S}^\prime$. When the transmission of the images from all perspectives is completed, we get the reconstructed multi-perspective images $\mathbf{\hat{I}}_\text{M}^\prime$.

\subsubsection{SKD-Based Model Training}
To ensure that ASCM learns to compress semantic information without compromising the quality of image reconstruction, we implement an SKD-based training approach. As illustrated in Fig. \ref{fig:ASCM2}, we consider the ASCM without semantic compression (i.e., there is no mask output head in the semantic encoder) as the ``teacher", while the ASCM with semantic compression (i.e., there is a mask output head in the semantic encoder) serves as the ``student". The task loss for the teacher can be expressed as follows:
\begin{equation}\label{eq:ASCM5}
	\mathcal{L}_\text{task}^\text{tech}= \mathcal{L}_\text{MSE}(\mathbf{I}_\text{S}^\prime,\mathbf{\hat{I}}_\text{S}^{\prime,+}),
\end{equation}
where $\mathbf{\hat{I}}_\text{S}^{\prime,+}$ is the reconstructed image without semantic compression in the transmission; $\mathcal{L}_\text{MSE}(\cdot)$ represents the mean square error. Hence, $\mathcal{L}_\text{task}^\text{tech}$ represents the reconstruction loss of the ASCM with complete semantic information and provides task-speciﬁc supervision for the teacher. Similarly, the task loss for the student can be given by:
\begin{equation}\label{eq:ASCM6}
	\mathcal{L}_\text{task}^\text{stu}= \mathcal{L}_\text{MSE}(\mathbf{I}_\text{S}^\prime,\mathbf{\hat{I}}_\text{S}^{\prime,-}),
\end{equation}
where $\mathbf{\hat{I}}_\text{S}^{\prime,-}$ is the reconstructed image with semantic compression in the transmission. 

During training, the teacher transfers the learned knowledge to the student, thus directing the student to learn well. This process can be represented as:
\begin{equation}\label{eq:ASCM7}
	\mathcal{L}_\text{KD}=\frac{\operatorname{KL}(\hat{\mathbf{E}}_s^{+},\hat{\mathbf{E}}_s^{-})}{\mathcal{L}_\text{task}^\text{tech}+\mathcal{L}_\text{task}^\text{stu}},
\end{equation}
where ${\operatorname{KL}}(\cdot)$ means the Kullback–Leibler divergence, i.e., ${\rm{KL}}(\mathbf{a},\mathbf{b})=-\sum_{i} \mathbf{a}_{i} \log \left(\mathbf{b}_{i} / \mathbf{a}_{i}\right)$; $\hat{\mathbf{E}}_s^{+}$ represents the reconstructed semantic information with the uncompressed semantics $\mathbf{E}_s^+$; $\hat{\mathbf{E}}_s^{-}$ is the reconstructed semantics corresponding to the compression semantics $\mathbf{E}_s^{-}$. $\mathcal{L}_\text{KD}$ is proposed to minimize the difference between the two reconstructed semantic information. Here, $\hat{\mathbf{E}}_s^{-}$ is expected to as possible as close to $\hat{\mathbf{E}}_s^{+}$. The $\mathcal{L}_\text{task}^\text{tech}+\mathcal{L}_\text{task}^\text{stu}$ is used to adjust the $\operatorname{KL}(\hat{\mathbf{E}}_s^{+},\hat{\mathbf{E}}_s^{-})$, adaptively.

By employing the SGD algorithm to minimize the aforementioned loss functions, we can effectively update the parameters of ASCM. Assuming that the number of training epochs is $J$, we summarize the ASCM in \textbf{Algorithm \ref{alg:ASCM}}.
 
\begin{algorithm}
	\caption{ASCM}
	\label{alg:ASCM}
	\begin{algorithmic}[1]
		\REQUIRE $\mathbf{I}_\text{S}^\prime$, $J$.
		\ENSURE $\mathbf{\vartheta}$, $\mathbf{\alpha}$, $\mathbf{\beta}$, $\mathbf{\delta}$.
		\STATE{\textbf{Training Stage}}
		\FOR{each epoch in $J$}
		\STATE{Calucate $\mathcal{L}_\text{task}^\text{tech}$ according to Eq. (\ref{eq:ASCM5})}.
		\STATE{Update the teacher model by minimizing $\mathcal{L}_\text{task}^\text{tech}$ with SGD optimizer.}
		\STATE{Calucate $\mathcal{L}_\text{task}^\text{stu}$ according to Eq. (\ref{eq:ASCM6}).}
		\STATE{Calucate $\mathcal{L}_\text{KD}$ according to Eq. (\ref{eq:ASCM7}).}
		\STATE{Update the student model (ASCM) by minimizing $\mathcal{L}_\text{stu} + \mathcal{L}_\text{KD}$ with SGD optimizer.}
		\ENDFOR
		\STATE{\textbf{Inference Stage}}
		\STATE{Obtain the emitted signals $\mathbf{X}$ by semantic and channel encoding according to Eq. (\ref{eq:Shi1}).}
		\STATE{Obtain the received signals $\mathbf{Y}$ according to Eq. (\ref{eq:Shi2}).}
		\STATE{With the obtained CSI $\hat{\mathbf{H}}_\text{ref}$ by GDCE, set $\mathbf{Y}=\mathbf{Y}/\hat{\mathbf{H}}_\text{ref}$.}
		\STATE{Obtain the recovered 3D scenario $\hat{\chi}^\text{3D}$ according to Eq. (\ref{eq:Shi3}).}
	\end{algorithmic}
\end{algorithm}

The benefits of ASCM can be encapsulated as follows: (1) We design two output heads in the semantic encoder to extract semantic features while performing semantic compression, eliminating the extra feature-level semantic information and reducing the communication cost; (2) We employ an SKD-based training method to ensure that the performance of ASCM with semantic compression approximates as closely as possible to that achieved with complete semantics, thus effectively maintaining the accuracy of the semantics.

\subsection{GDCE}
This subsection details the implementation of GDCE.
According to Eqs. (\ref{eq:Shi2}) and (\ref{eq:ASCM4}), we can obtain the following formula:
\begin{equation}\label{eq:GDCE1}
	\hat{\mathbf{E}_{s}^{-}} =
	\sigma\left(\mathbf{HX} \mathbf{w}_{r}+\mathbf{N}\mathbf{w}_{r}+\mathbf{b}_{r}\right).
\end{equation}

In Eq. (\ref{eq:GDCE1}), the task of $\mathbf{w}_r$ is how to accurately recover the transmitted signals affected by the channel effects. This increases the training burden and limits the expressive capability of the network. Furthermore, errors introduced by channel effects propagate to subsequent layers of the ASCM decoder, which further compounds the problem.
Moreover, since $\mathbf{H}$ is untrainable and random, it introduces perturbations during weight updating. This means that weight updating occurs with higher variance. Therefore, even though minimizing the loss functions in Eqs. (\ref{eq:ASCM5}) - (\ref{eq:ASCM7}) can optimize ASCM to some extent, the fading channel still contaminates gradients during back-propagation and constrains semantic representation during forward-propagation \cite{xie2020lite}.

As a solution, we enhance the recovery of signals by leveraging CSI between the transmitter and receiver. When $\mathbf{H}$ is known, according to Eq. (\ref{eq:Shi2}), we can calculate the received signals as follows:
{\color{black}
\begin{equation}\label{eq:GDCE2}
\begin{aligned}
	& \mathbf{Y}=\mathbf{HX}+\mathbf{N} \\
	\Rightarrow & {\mathbf{H}^{-1}}\mathbf{Y}=\mathbf{X}+\mathbf{H}^{-1}\mathbf{N} \\
	\Rightarrow & {(\mathbf{H}^{H}\mathbf{H})}^{-1}\mathbf{Y}={(\mathbf{H}^{H})}^{-1}\mathbf{X}+{(\mathbf{H}^{H}\mathbf{H})}^{-1}\mathbf{N} \\
	\Rightarrow & \tilde{\mathbf{Y}}={(\mathbf{H}^{H}\mathbf{H})}^{-1}\mathbf{H}^{H}\mathbf{Y}=\mathbf{X}+{(\mathbf{H}^{H}\mathbf{H})}^{-1}\mathbf{H}^{H}\mathbf{N}. \\
\end{aligned}
\end{equation}
}
We denote $\tilde{\mathbf{N}}=\left(\mathbf{H}^{H} \mathbf{H}\right)^{-1} \mathbf{H}^{H} \mathbf{N}$. Compared to Eq. (\ref{eq:Shi2}), in Eq. (\ref{eq:GDCE2}), the channel effect transitions from multiplicative noise to additive noise $\tilde{\mathbf{N}}$. This shift enables stable back-propagation and enhances the representational capability of the ASCM.

\begin{figure}[htbp]
	\centering
	\includegraphics[width=8.5cm]{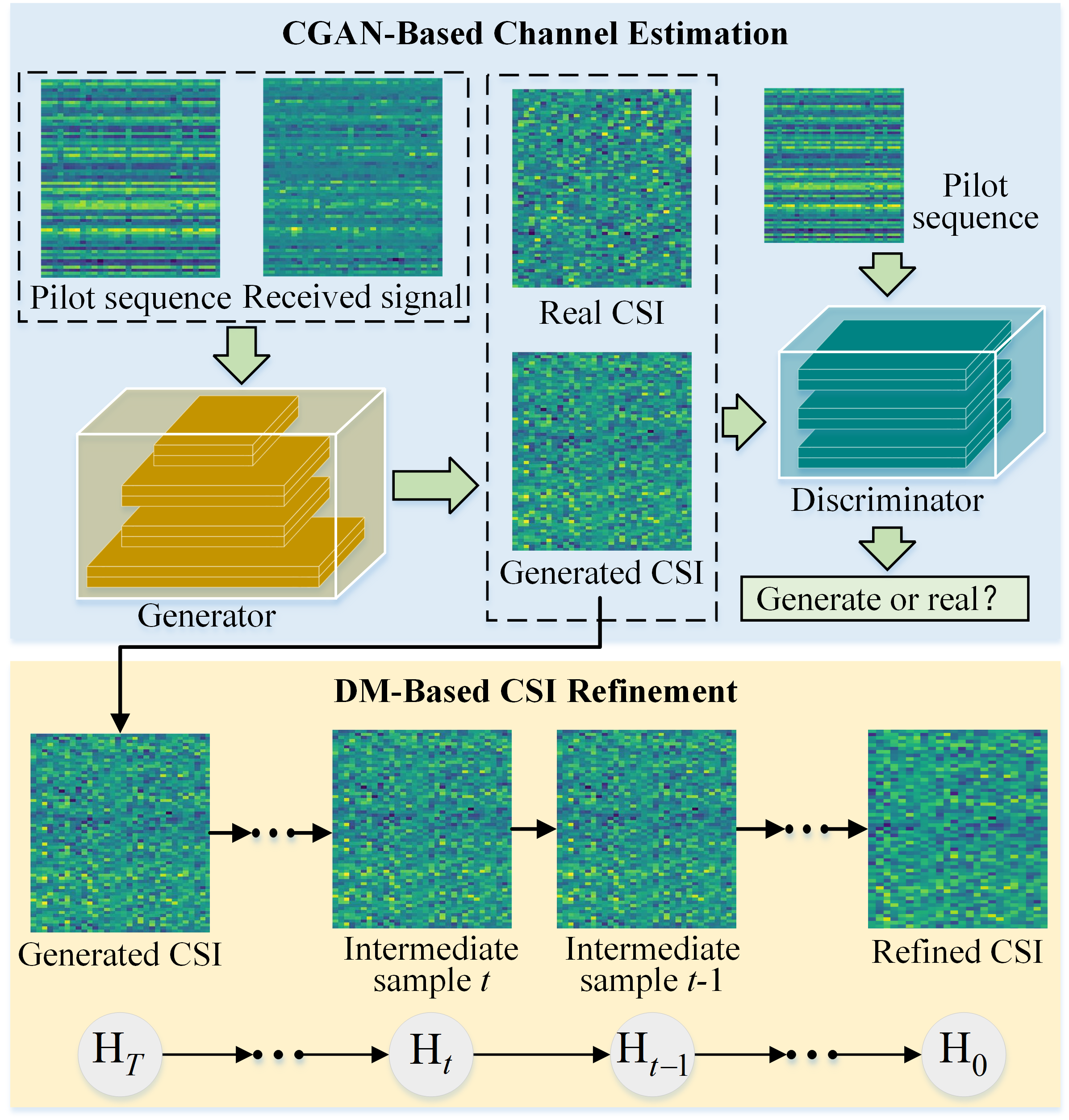}
	\caption{The illustration of GDCE.}
	\label{fig:GDCE}
\end{figure}

To accurately obtain $\mathbf{H}$ between the transmitter and receiver, we propose the GDCE method in this paper. As illustrated in Fig. \ref{fig:GDCE}, GDCE mainly comprises the following two components:
\subsubsection{CGAN-Based Channel Estimation}
Firstly, we can denote the received signal $\mathbf{Y}$, the pilot sequence $\mathbf{\Theta}$, and channel matrix $\mathbf{H}$ as dual-channel images \cite{dong2020channel}. The two channels represent the real and imaginary parts of a complex matrix. From this perspective, we can redefine the problem of channel estimation as an image-to-image generative task.

Secondly, within the CGAN, the generator uses the pilot sequence $\mathbf{\Theta}$ and the received signal $\mathbf{Y}$ as input to estimate $\mathbf{H}$, denoted as $\hat{\mathbf{H}}$. In this case, the pilot sequence $\mathbf{\Theta}$ is used as the condition. The goal of the discriminator is to distinguish between the real and generated CSI, with the pilot sequence $\mathbf{\Theta}$ serving as the condition.

Thirdly, CGAN aims for the generator to synthesize highly realistic CSI that can deceive the discriminator. Correspondingly, the discriminator tries to improve its discernment capabilities so it is not easily tricked. To optimize this process, we apply a least-squares GAN loss function \cite{mao2017least} that incorporates conditional information into consideration. The loss functions for both discriminator $D$ and generator $G$ are expressed below:
\begin{equation}\label{eq:GDCE3}
	\min_{D} \mathcal{L}_{\text{d}}= \mathbb{E}\left[(D(\mathbf{H}|\mathbf{\Theta})-1)^{2}\right]+ \mathbb{E}\left[(D(G(\mathbf{Y}|\mathbf{\Theta}))+1)^{2}\right],
\end{equation}
\begin{equation}\label{eq:GDCE4}
	\min_{G} \mathcal{L}_{\text{g}}= \mathbb{E}\left[(D(G(\mathbf{Y}|\mathbf{\Theta})))^{2}\right].
\end{equation}

Additionally, to ensure the CSI generated by the generator closely aligns with the real CSI at a pixel level, we incorporate an L1 loss into the loss function of the generator:
\begin{equation}\label{eq:GDCE5}
	\mathcal{L}_{\text{L1}}= \mathbb{E}\left[\rVert(D(\mathbf{H}|\mathbf{\Theta})-\mathbf{H})\rVert\right],
\end{equation}
where $\rVert\cdot\rVert$ denotes the L1 loss function. 

\subsubsection{DM-based CSI Refinement}
{\color{black}
Due to the prior constraints on CGAN, the generated CSI images often lack fine details. Therefore, an additional DM is employed to enhance the predicted details of CSI $\hat{\mathbf{H}}$. Therefore, we employ a DM to further refine the CSI and thereby enhance its accuracy.

The training of DM encompasses two procedures: a forward process (diffusion process) and a reverse process (generation process). The forward process involves gradually adding Gaussian noise $\mathbf{z}$ to the predicted CSI $\hat{\mathbf{H}}$ until it transforms into pure noise $\hat{\mathbf{H}}_T$. Each step adheres to the following Markov chain:
\begin{equation}\label{eq:GDCE7}
q(\hat{\mathbf{H}}_t|\hat{\mathbf{H}}_{t-1}) = \mathcal{N}(\hat{\mathbf{H}}_t; \sqrt{\alpha_t} \hat{\mathbf{H}}_{t-1}, (1-\alpha_t)\mathbf{I}),
\end{equation}
where $\alpha_t = 1 - \beta_t$ is a predefined decreasing sequence that satisfies $\beta_1 < \beta_2 < \cdots < \beta_T$; $\hat{\mathbf{H}}_t$ is the intermediate sample at the $t$-th step. This ensures that each step has the same noise diffusion amplitude.

The reverse process is the procedure of progressively reconstructing the original image $\mathbf{H}$ from the pure noise $\hat{\mathbf{H}}_T$. Each step involves a neural network predicting a denoising function $p(\hat{\mathbf{H}}_{t-1}|\hat{\mathbf{H}}_t)$, designed to fit the true posterior distribution $q(\hat{\mathbf{H}}_{t-1}|\hat{\mathbf{H}}_t, \hat{\mathbf{H}}_0)$. 

We regard the refined CSI is denoted as $\hat{\mathbf{H}}_\text{ref}$, while the mean of the denoising function $p(\hat{\mathbf{H}}_{t-1}|\hat{\mathbf{H}}_t)$ is represented as $\mu_{\theta}(\hat{\mathbf{H}}_t, t)$, where $\theta$ represents the parameters of DM. $\mathbf{z}$ represents Gaussian noise. 
We assume the training epoch of GDCE is $O$.
Finally, we summarize the training and inference stages of GDCE in \textbf{Algorithm \ref{alg:GDCE}}.
}

The GDCE has two benefits as follows: 
(1) We employ a CGAN to estimate the CSI based on pilot sequences and the received signals. This process achieves the transition of channel effects from multiplicative noise to additive noise and thus help the recovery of signals in the receiver. (2) We utilize a DM to refine the CSI produced by CGAN, with an aim to enhance CSI image details. This further improves the accuracy of the obtained CSI.
\begin{algorithm}[H]
	{\color{black}
	\caption{GDCE}
	\label{alg:GDCE}
	\begin{algorithmic}[1]
		\REQUIRE $\mathbf{H}$, $\mathbf{\Theta}$, $O$, $\theta$.
		\ENSURE $G$, $D$, $\theta$.
		\STATE{\textbf{Training Stage}}
		\STATE{Initialize $G$ $D$ and $\theta$.} 
		\FOR{$n = 1, 2, \cdots, O$}
		\STATE{Calculate the loss of the discriminator according to Eq. (\ref{eq:GDCE3}) and update $D$ via SGD.}
		\STATE{Calculate the loss of the generator according to Eq. (\ref{eq:GDCE4}) and Eq. (\ref{eq:GDCE5}) and update $G$ via SGD.}
		\STATE{Obtain predicted CSI $\hat{\mathbf{H}}=G(\mathbf{Y}|\mathbf{\Theta})$}. 
		\STATE{Randomly sample a time step $t$ from $\{1,2,\dots,T\}$.}
		\STATE{Obtain $q=q(\hat{\mathbf{H}}_{t-1}|\hat{\mathbf{H}}_t, \hat{\mathbf{H}}_0)$ according to Eq. (\ref{eq:GDCE7}).}
		\STATE{Calculate the loss function $\mathcal{L}(\theta) = \|q - \mu_\theta(\hat{\mathbf{H}}_t, t)\|^2$ and update $\theta$ via SGD.}
		\ENDFOR
		\STATE{\textbf{Inference Stage}}
		\STATE{Obtain predicted CSI $\hat{\mathbf{H}}=G(\mathbf{Y}|\mathbf{\Theta})$}. 
		\STATE{$\hat{\mathbf{H}}_T = \hat{\mathbf{H}}$.}
		\FOR{$t = T, T-1, \cdots, 1$}
		\STATE{$\mathbf{z} \sim \mathcal{N}(0, 1)$.}
		\STATE{$\hat{\mathbf{H}}_{t-1} = \mu(\hat{\mathbf{H}}_t, t) + \sqrt{\beta_t} \mathbf{z}$.}
		\ENDFOR
		\STATE{$\hat{\mathbf{H}}_\text{ref} = \hat{\mathbf{H}}_0 = \hat{\mathbf{H}}_1 / \sqrt{\alpha_1}$.}
	\end{algorithmic}
}
\end{algorithm}

\section{Numerical Results}
To evaluate the effectiveness of the proposed methods, we conducted a series of simulations. These simulations are performed on a server equipped with an Intel Xeon CPU (2.4 GHz, 128 GB RAM) and an NVIDIA A100 GPU (80 GB SGRAM), using the PyTorch framework to implement the SC schemes.

In the proposed GAM-3DSC system, we directly apply the pre-trained weights of the NeRF and SAM from \cite{cen2023segment}, hence there is no need to retrain the 3DSE. The ASCM and GDCE can be trained independently and then used in conjunction. 
Next, we detail the simulation settings and the evaluation results of semantic extraction based on 3DSE, image transmission based on ASCM, channel estimation based on GDCE, and 3D transmission based on GAM-3DSC.

\subsection{Evaluation for Semantic Extraction Based on 3DSE}
This subsection showcases the semantic extraction results of the proposed 3DSE scheme. 

\subsubsection{Simulation Settings}
We implement 3DSE on the LLFF and mip-NeRF360 datasets \cite{cen2023segment}, comprised of various images of 3D scenarios captured from different angles.
Subsequently, we employ two forms of prompts (i.e., points and text) to evaluate the performance of 3D semantic extraction.

\subsubsection{Evaluation Results}
\begin{figure*}[htbp]
	\centering
	\includegraphics[width=18cm]{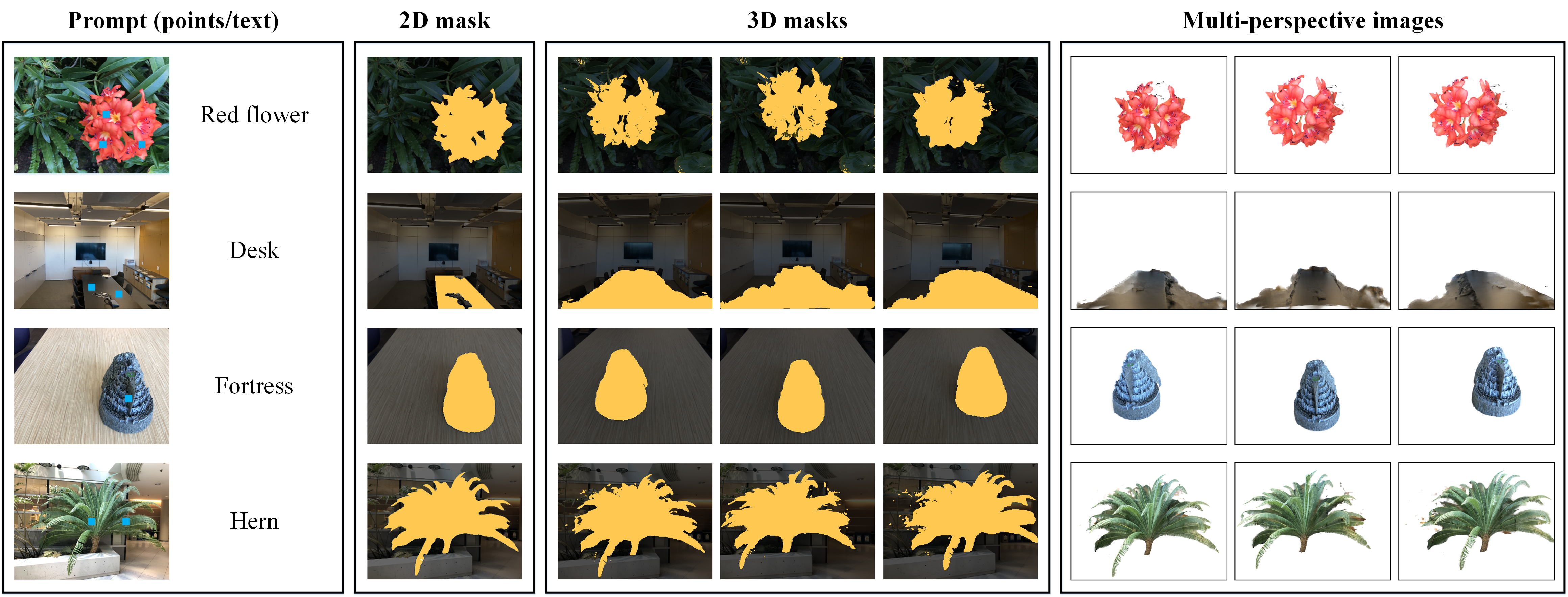}
	\caption{Partial results of semantic extraction by 3DSE.}
	\label{fig:3DSE_exp}
\end{figure*}

Fig. \ref{fig:3DSE_exp} presents partial results of semantic extraction by 3DSE. The first box on the left demonstrates two forms of prompts: one is selecting the 3D object of interest using several points, and the other is providing the name of the 3D object in text form. The second box displays the 2D mask corresponding to the user's prompts. The third box showcases the 3D mask derived from mask inverse rendering, illustrating that 3DSE can obtain masks of all multi-perspective images based solely on a mask from a single-perspective image. The final box presents the multi-perspective images of the selected 3D object, which constitute the transmitted image data.

In summary, the results demonstrate that the implemented 3DSE can accurately extract the intended 3D object (i.e., key semantics) based on the user prompts. The utilization of NeRF and SAM can provide robust capabilities for 3D object processing and image segmentation.

\subsection{Evaluation for Image Transmission Based on ASCM}
This subsection is intended to present the evaluation results of the proposed ASCM scheme during its training phase. 

\subsubsection{Simulation Settings}
Firstly, we utilize the multi-perspective images extracted from the LLFF, mip-NeRF360, and LERF datasets by 3DSE to train the ASCM.

Secondly, to highlight the advantages of the transformer architecture in ASCM, we compare it with several other architecture-based SC methods. These include Convolutional Autoencoder-based SC (CAE-SC), Variational Autoencoder-based SC (VAE-SC), and Vision Transformer-based SC (ViT-SC). Apart from ASCM, all other methods transmit complete semantic information and employ the SGD algorithm for updates. Except for ASCM, the rest models adopt the Deconvolution Networks (DCNNs) \cite{noh2015learning} architecture as the semantic decoder. For a more comprehensive understanding, we have summarized the architectures of these methods in Table \ref{tab:ASCM}.

\begin{table}[htpb]
	\centering\makegapedcells
	\caption{Summarization of SC schemes.}
	\label{tab:ASCM}
\begin{tabular}{|p{45pt}<{\centering}|cccc|}
	\hline
	& \multicolumn{1}{c|}{ASCM}                & \multicolumn{1}{c|}{CAE-SC} & \multicolumn{1}{c|}{VAE-SC} & ViT-SC      \\ \hline
	Semantic encoder          & \multicolumn{1}{c|}{Transformer}         & \multicolumn{1}{c|}{CNN}    & \multicolumn{1}{c|}{MLP}    & Transformer \\ \hline
	Channel encoder/decoder   & \multicolumn{4}{c|}{MLP}                                                                                           \\ \hline
	Semantic decoder          &
	\multicolumn{1}{c|}{Transformer}         &
	\multicolumn{3}{c|}{DCNN}                                                                                          \\ \hline
	Semantic compression      & \multicolumn{1}{c|}{Yes}                 & \multicolumn{3}{c|}{No}                                                 \\ \hline
	Size of data transmission & \multicolumn{1}{c|}{\textbf{4,326 bits}} & \multicolumn{3}{c|}{21,632 bits}                                        \\ \hline
\end{tabular}
\end{table}
\subsubsection{Evaluation Results}
\begin{figure*}[htbp]
	\centering
	\includegraphics[width=17cm]{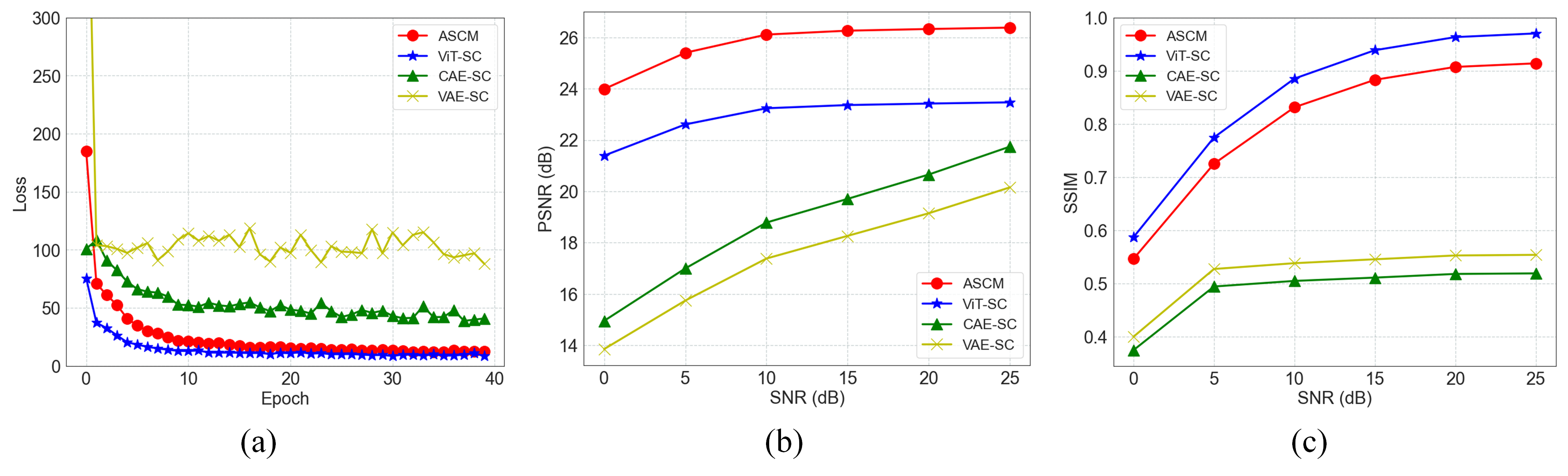}
	\caption{Evaluation results of different SC models under AWGN channel. (a) Loss versus epoch. (b) PSNR versus SNR. (c) SSIM versus SNR.}
	\label{fig:ASCM_exp1}
\end{figure*}
\begin{figure*}[htbp]
	\centering
	\includegraphics[width=17cm]{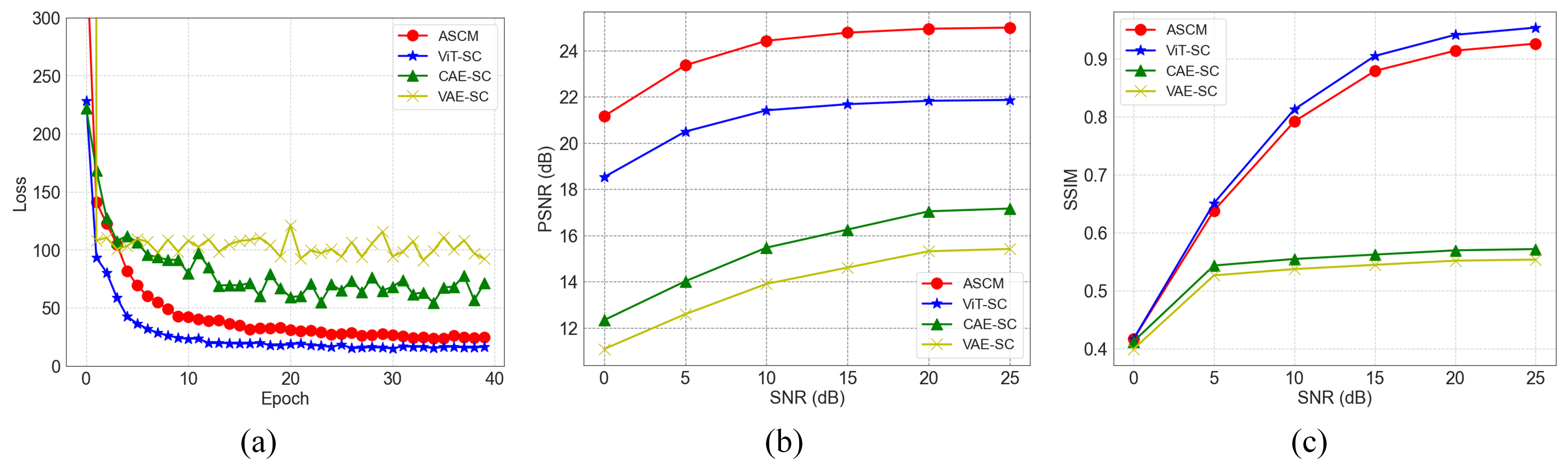}
	\caption{Evaluation results of different SC models under the fading channel. (a) Loss versus epoch. (b) PSNR versus SNR. (c) SSIM versus SNR.}
	\label{fig:ASCM_exp2}
\end{figure*}

Thirdly, the evaluation metrics include loss, Peak Signal-to-Noise Ratio (PSNR), and Structural Similarity Index Measure (SSIM). PSNR serves as a measure of the quality of a reconstructed or compressed image. It is typically expressed in decibels, with higher values denoting superior image quality. The definition of PSNR is as follows:
\begin{equation}
	\mathrm{PSNR}(\mathbf{I}_\text{S}^\prime,\mathbf{\hat{I}}_\text{S}^\prime) = 10 \cdot \log_{10} \left( \frac{\mathrm{MAX}_I^2}{\mathrm{MSE}(\mathbf{I}_\text{S}^\prime,\mathbf{\hat{I}}_\text{S}^\prime)} \right),
\end{equation}
where $\mathrm{MAX}_I$ denotes the maximum possible pixel value of the image, which is typically 255 for an 8-bit image. $\mathrm{MSE}(\cdot)$ represents the average squared difference between the original image $\mathbf{I}_\text{S}^\prime$ and the reconstructed image $\mathbf{\hat{I}}_\text{S}^\prime$.
Similarly, SSIM is a metric that gauges the perceived similarity between two images, factoring in three key components - luminance, contrast, and structure. The definition of SSIM is outlined as follows:
\begin{equation}
	\mathrm{SSIM}(\mathbf{I}_\text{S}^\prime,\mathbf{\hat{I}}_\text{S}^\prime) = \frac{(2\varphi_{\mathbf{I}_\text{S}^\prime}\varphi_{\mathbf{\hat{I}}_\text{S}^\prime} + c_1)(2\phi_{\mathbf{I}_\text{S}^\prime\mathbf{\hat{I}}_\text{S}^\prime} + c_2)}{(\varphi_{\mathbf{I}_\text{S}^\prime}^2 + \varphi_{\mathbf{\hat{I}}_\text{S}^\prime}^2 + c_1)(\phi_{\mathbf{I}_\text{S}^\prime}^2 + \phi_{\mathbf{\hat{I}}_\text{S}^\prime}^2 + c_2)},
\end{equation}
where $\varphi_{\mathbf{I}_\text{S}^\prime}$ and $\varphi_{\mathbf{\hat{I}}_\text{S}^\prime}$ are their means; $\phi_{\mathbf{I}_\text{S}^\prime}^2$ and $\phi_{\mathbf{\hat{I}}_\text{S}^\prime}^2$ are their variances; $\phi_{\mathbf{I}_\text{S}^\prime\mathbf{\hat{I}}_\text{S}^\prime}$ is their covariance; $c_1$ and $c_2$ are two constants used to avoid division by zero.

Finally, we consider two types of channels, namely AWGN and fading  (Rician channel is used in this paper) channels in the data transmission. Signal-to-Noise Ratio (SNR) range is from 0 dB to 25 dB. The training epoch is set to 40.

Fig. \ref{fig:ASCM_exp1} and \ref{fig:ASCM_exp2} illustrate the evaluation results of various SC models under AWGN and fading channels, respectively. 
Initially, as depicted in Fig. \ref{fig:ASCM_exp1}(a) and Fig. \ref{fig:ASCM_exp2}(a), the final convergence results of the proposed ASCM outperform the CAE-SC and VAE-SC schemes, but are slightly inferior to the ViT-SC scheme. 
Subsequently, Fig. \ref{fig:ASCM_exp1}(b) and Fig. \ref{fig:ASCM_exp2}(b) demonstrate that the SC models attain a comparable PSNR under both channels when the SNR is sufficiently high. It is observable that the CAE-SC and VAE-SC schemes perform poorly, while the ASCM and ViT-SC schemes perform better. Additionally, the performance gap between the ASCM and ViT-SC schemes is narrower on the fading channel than on the AWGN channel.
Lastly, Fig. \ref{fig:ASCM_exp1}(c) and Fig. \ref{fig:ASCM_exp2}(c) indicate that the various SC schemes obtain similar results to those in Fig. \ref{fig:ASCM_exp1}(b) and Fig. \ref{fig:ASCM_exp2}(b) when the metric is SSIM. This illustrates that the ASCM can ensure consistency of the sent and received images at the pixel level.

The superior performance of the ViT-SC and ASCM schemes results from the benefits of the transformer architecture, which extracts more precise semantic information compared to the CNN and MLP architectures. Since the ViT-SC scheme does not perform semantic compression, it slightly outperforms ASCM. However, as Table \ref{tab:ASCM} shows, ASCM reduces the amount of data to be transferred by 80\%. This suggests that while ASCM sacrifices some model accuracy, it significantly reduces transmission energy consumption. The SKD-based training method plays an important role in this, as it enables ASCM to reduce the size of transmitted semantics while keeping the most critical semantic information.

\subsection{Evaluation for Channel Estimation Based on GDCE}
This subsection showcases the evaluation results of the proposed GDCE scheme, specifically during its training phase. 

\subsubsection{Simulation Settings}
Firstly, we simulate a scenario where two devices communicate at a carrier frequency of 24.2 GHz within the 28 GHz millimeter wave band. Following this, we employ the Rician fading channel model as the physical channel to simulate the path loss, utilizing the path loss model in the urban microcell street scenario as per the 3GPP TR 38.901 standards.

Secondly, we conduct the ASCM in a simulated wireless environment, where we gather the output of the channel encoder in ASCM as the pilot sequences and the input of the channel decoder as the received signals. For training the CGAN, we collect 1800 pairs of samples in total.

Then, we compare the proposed GDCE to both traditional channel estimation methods and DL methods. The traditional channel estimation contenders include the Least Squares (LS) channel estimator, Orthogonal Matching Pursuit (OMP), Approximate Message Passing (AMP), and the Minimum Mean Square Error (MMSE) channel estimator. For the DL contenders, we select different architectures-based channel estimators, including U-net, MLP, and CGAN.

Finally, we use the Normalized Mean Square Error (NMSE) as the evaluation metric, which can assess the error between the predicted and actual results. The calculation formula for NMSE is as follows:
\begin{equation}
	\operatorname{NMSE} = 10\log_{10}\{\frac{ (\mathbf{H}- \hat{\mathbf{H}})^2}{\operatorname{Var}(\mathbf{H})}\},
\end{equation}
where $\operatorname{Var}(\mathbf{H})$ represents the variance of $\mathbf{H}$. The smaller the value of NMSE, the closer the predicted results are to the real results.

\subsubsection{Evaluation Results}
\begin{figure}[htbp]
	\centering
	\includegraphics[width=8.5cm]{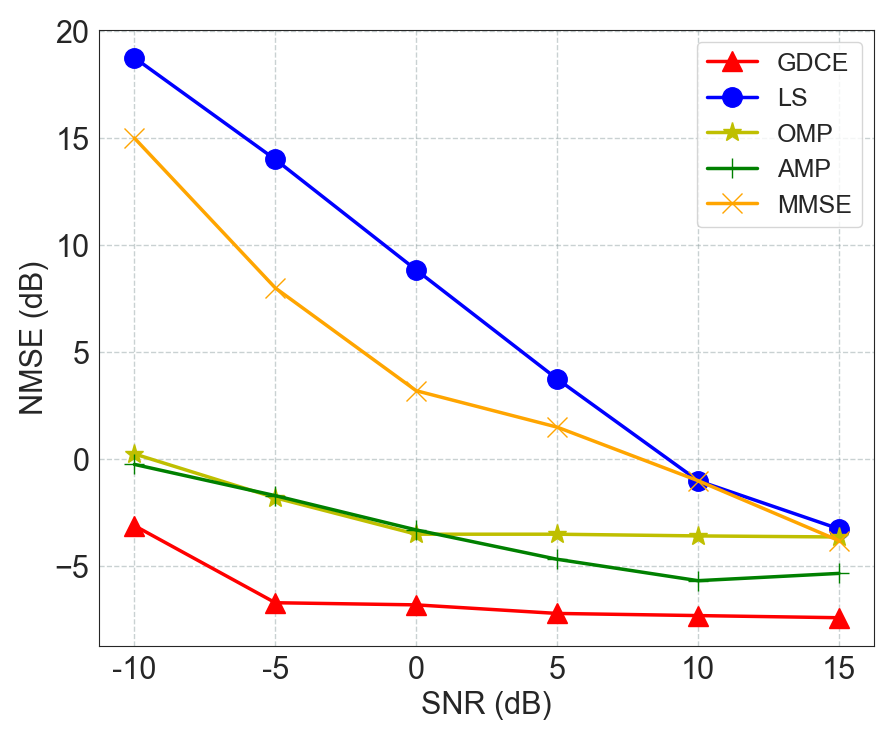}
	\caption{NMSE of different schemes under different SNR.}
	\label{fig:GDCE2}
\end{figure}

\begin{figure}[htbp]
	\centering
	\includegraphics[width=8.5cm]{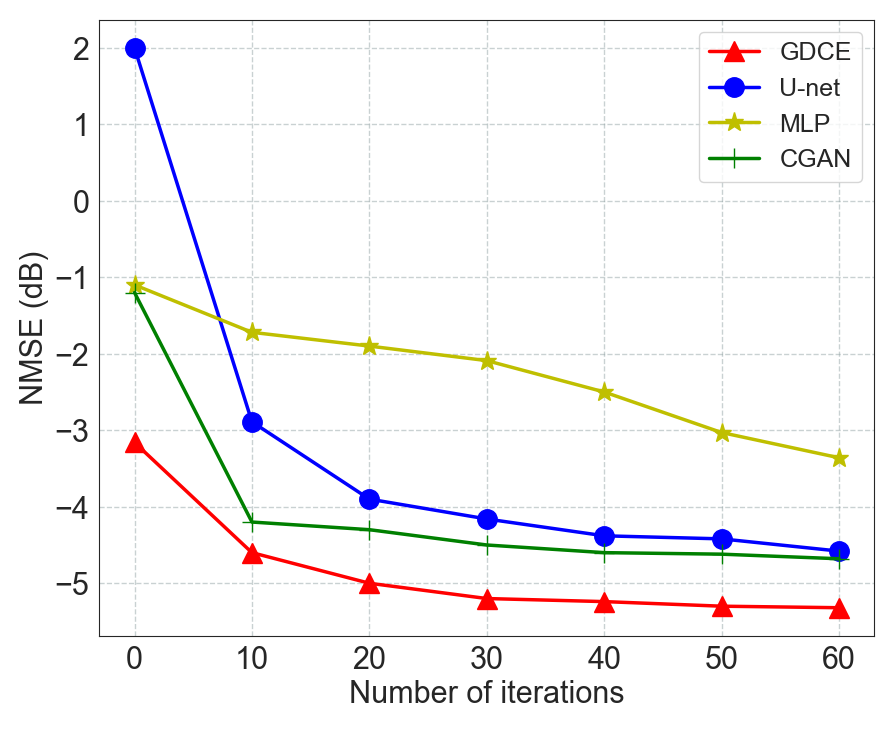}
	\caption{NMSE of different schemes under different iterations.}
	\label{fig:GDCE3}
\end{figure}

Fig. \ref{fig:GDCE2} depicts the NMSE of different channel estimation schemes in different SNRs. As SNR improves, so does the performance of each scheme. The proposed GDCE consistently achieves the lowest NMSE across all SNRs, while the LS channel estimator continues to perform the worst.
Fig. \ref{fig:GDCE3} demonstrates the NMSE of different DL-based channel estimators as the iterations increase. It is apparent that GDCE and CGAN outperform the other methods, highlighting the superiority of the CGAN architecture.


The superior performance of GDCE stems from two factors. Firstly, the CGAN architecture allows us to achieve more realistic CSI even with limited pilots. Secondly, the DM-based refinement strategy removes noise from the generation process of CGAN, thereby improving the accuracy of channel estimation.
In conclusion, the effective utilization of the CGAN architecture and the DM-based refinement strategy accounts for the exceptional performance of the GDCE, collectively enhancing channel estimation accuracy.

\subsection{Evaluation for 3D transmission based on GAM-3DSC}
This subsection aims to evaluate the performance of 3D transmission based on the proposed GAM-3DSC system.

\subsubsection{Simulation Settings}
We employ the LLFF and mip-NeRF360 as our experimental datasets. Specifically, we use the GAM-3DSC to transmit data and recover the 3D scenario. We then assess the difference between the raw and recovered 3D scenarios. As our primary concern is the semantics loss of the selected 3D object, we use the 3DSE to process the raw 3D scenario and generate a new 3D scenario containing only the selected 3D object. We then compare this processed 3D scenario with the recovered one. We utilize the following two evaluation methods:
\begin{itemize}
	\item \emph{Pixel-level evaluation}:
	We adopt the evaluation method proposed in \cite{kirillov2023segment}, in which we first obtain multiple images from the same perspective in the processed 3D and recovered scenarios respectively. Furthermore, we compare the pixel-level differences between images from the original 3D scenario and those from the reconstructed 3D scenario, captured from the same perspective. The metrics we employ are PSNR and SSIM.
	
	\item \emph{Semantic-level evaluation}:
We use LAMs for semantic-level evaluation. Firstly, we adopt the BLIP \cite{li2022blip}, a large visual language model that unifies the tasks of visual language understanding and generation, to transform the multi-perceptive images into text. Then, we adopt the large language model such as BERT \cite{devlin2018bert} to obtain the embeddings of these texts. Finally, we compare the difference between the embeddings by BLEU score and cosine similarity \cite{mrinalini2022sbsim}.
\end{itemize}   

\subsubsection{Evaluation Results}

\begin{figure}[htbp]
	\centering
	\includegraphics[width=9cm]{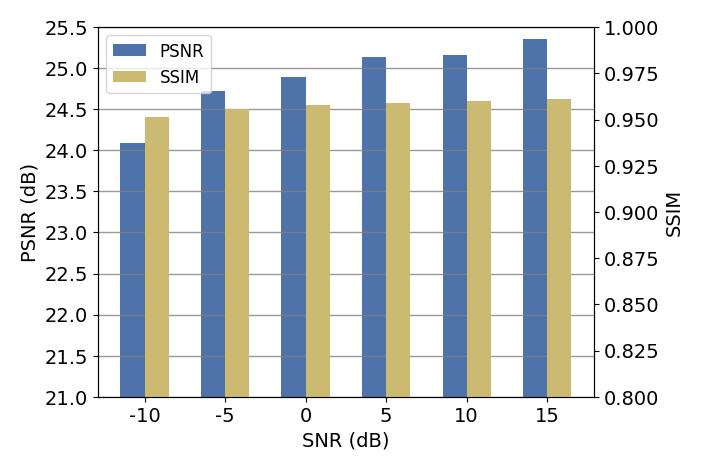}
	\caption{Pix-level evaluation results under different SNRs.}
	\label{fig:LAM3DSC1}
\end{figure}
\begin{figure}[htbp]
	\centering
	\includegraphics[width=9cm]{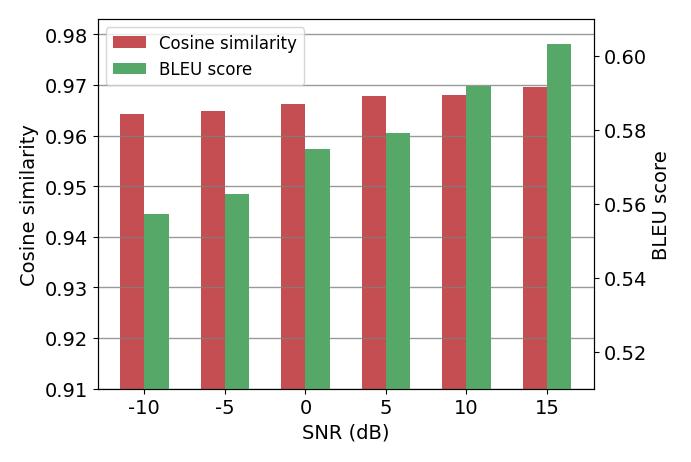}
	\caption{Semantic-level evaluation results under different SNRs.}
	\label{fig:LAM3DSC2}
\end{figure}

Fig. \ref{fig:LAM3DSC1} shows the pix-level evaluation results, in which we can see the performance of the GAM-3DSC in terms of the PSNR and SSIM is increased with the improvement of SNR. However, the change of the SNR and SSIM is small between the low and high SNR, which reflects the anti-interference ability against channel noise.
Furthermore, the proposed GAM-3DSC system can achieve a PSNR value of approximately 25 dB and an SSIM value of about 0.95. This indicates that, despite potential variations in pixel values arising from differences in brightness, contrast, or color within the reconstructed image, the structural similarity between the original and reconstructed images is substantial, ensuring similarity between images.
Fig. \ref{fig:LAM3DSC2} showcases the semantic-level evaluation results. We can see similar results in Fig. \ref{fig:LAM3DSC1}. 
The BLEU score can reach a maximum of $0.61$ and the cosine similarity can reach a maximum of $0.97$. This shows that although the contrasting texts vary in word composition, their underlying semantics are remarkably similar. These evaluation results confirm that the proposed GAM-3DSC can transmit the 3D object while preserving semantic consistency.

\section{Conclusions}
In this paper, we propose a GAM-3DSC system for addressing various challenges when implementing 3D SC.
We first introduce 3DSE which uses SAM and NeRF  to extract key semantics from a 3D scenario based on user requirements. The key semantics are represented as multi-perspective images of the selected 3D object.
We then propose ASCM to transmit these multi-perspective images, in which a semantic encoder with dual output heads performs semantic encoding and masks redundant information in the transmitted semantics.
Next, we apply GDCE to estimate and refine the CSI of the physical channel, thus contributing to the recovery of the signals in the receiver.
Finally, simulation results showcase the effectiveness of the proposed GAM-3DSC system.


\bibliographystyle{ieeetran}
\bibliography{bare_jrnl_bobo}
\newpage
\end{document}